\documentclass[12pt]{article}

\usepackage{graphicx}

\begin{document}

\begin{center}
{\bf Universe Inflation Based on Nonlinear Electrodynamics} \\
\vspace{5mm} S. I. Kruglov
\footnote{E-mail: serguei.krouglov@utoronto.ca}

\vspace{3mm}
\textit{Department of Physics, University of Toronto, \\60 St. Georges St.,
Toronto, ON M5S 1A7, Canada\\
Department of Chemical and Physical Sciences, University of Toronto,\\
3359 Mississauga Road North, Mississauga, ON L5L 1C6, Canada} \\
\vspace{5mm}
\end{center}

\begin{abstract}
A new model of nonlinear electrodynamics with dimensional parameters $\beta$ and $\gamma$ is proposed. The  principles of causality and unitarity are studied. We show that a singularity of the electric field at the origin of charges is absent and the maximum of the electric field in the center is $E(0)=1/\sqrt{\beta}$. The dual symmetry is broken in our model. Corrections to the Coulomb law as $r\rightarrow\infty$ are in the order of ${\cal O}(r^{-4})$. The source of the gravitation field and inflation of the universe is electromagnetic fields. It is supposed that the universe is filled by stochastic magnetic fields. We demonstrate that after the universe inflation the universe decelerates approaching the Minkowski spacetime. The singularities of the Ricci scalar, the Ricci tensor squared and the Kretschmann scalar are absent. We calculate the speed of sound. The spectral index, the tensor-to-scalar ratio, and the running of the spectral index, which approximately agree with the PLANK and WMAP data, are evaluated.
\end{abstract}

\section{Introduction}

By modifying general relativity (GR) ($F(R)$ gravity), one can explain the universe inflation. There are many different functions $F(R)$ that may explain inflation and, as a result, there are many modified gravity models \cite{Capozziello}, \cite{Nojiri}. At very strong electromagnetic fields, in early time of the universe evolution, quantum corrections have to be taken into consideration \cite{Jackson}, and Maxwell's electrodynamics becomes nonlinear electrodynamics (NED).  Firstly, Born and Infeld \cite{Born} proposed NED that removes a singularity of point-like charges and results in the finite value of self-energy. Quantum electrodynamics taking into account quantum corrections becomes NED \cite{Heisenberg}-\cite{Adler}.
 The gravitation field coupled to NED can describe inflation \cite{Garcia}-\cite{Kruglov4}
Here, we propose a new NED which, for weak fields, converts into Maxwell's electrodynamics and the correspondence principle holds. It will be shown, that in the framework of our NED coupled to the gravitational field, that the universe inflation occurs for the stochastic magnetic background.

It is known that there are the stochastic fluctuations of the electromagnetic field in a relativistic electron-positron plasma. Therefore plasma fluctuations can be the source of a stochastic magnetic field  \cite{Lemoine}, \cite{Lemoine1}. A primordial magnetic field is generated from thermal fluctuations in the pre-recombination plasma, and magnetic fluctuations are sustained by plasma before the epoch of Big Bang nucleosynthesis. The early universe was filled by a strong low-frequency random magnetic field during the early stage of the radiation-dominated era. Magnetic fields of the order of $B = 10^{-6}$ G exist on scales of a few Kpc in our galaxy and other spiral galaxies \cite{Kronberg}. Such magnetic fields possess the primordial origin and are explained by a mechanism transferring angular momentum energy into magnetic energy (the galactic dynamo theory). The galactic dynamo theory requires the existence of weak seed fields. For successful dynamo amplification a seed field of the order of $B= 10^{-19}$ G is required at the epoch of the galaxy formation. Seed magnetic fields may be generated by thermal fluctuations in the primordial plasma. Long wavelength fluctuations can reconnect and redistribute the magnetic energy over larger scales. A new scenario for the creation of galactic magnetic fields was proposed in \cite{Berezhiani}. It is worth noting that the origin of cosmic magnetism on the largest scales of galaxies, galaxy clusters and the general inter galactic medium is still an open problem \cite{Gaensler}.
In the following we consider the case when $E = 0$ because the electric field is screened by the charged primordial plasma, but the magnetic field is not screened  \cite{Lemoine1}. In accordance with the standard cosmological model, there is no asymmetry in the direction, and therefore $\langle B_i\rangle= 0$. Thus, the magnetic field does not induce the directional effects.

The structure of the paper is as follows. In Sect. 2 we propose a new model of NED with a dimensional parameters $\beta$ and $\gamma$. The causality and unitarity principles are studied. Field equations and their dual invariance are analyzed in Sect. 3 . We demonstrate that there is no singularity of the electric field in the center of the point-like charges and we find the maximum of the electric field. The corrections to Coulomb's law are obtained. The cosmology of the universe with stochastic magnetic fields is considered in Sect. 4. We obtain the energy density and pressure depending on the scale factor. It is shown that the singularity of the Ricci scalar is absent. In Sect. 5 the evolution of the universe is investigated. We find the dependence of the scale factor on the time. The speed of sound is calculated. The cosmological parameters, the spectral index $n_s$, the tensor-to-scalar ratio $r$, and the running of the spectral index $\alpha_s$, are evaluated in Sect. 6. We show that they are in approximate agreement with the PLANK and WMAP data. Section 7 is devoted to a conclusion.

We use the units with $c=\hbar=\varepsilon_0=\mu_0=1$ and the metric signature $\eta=\mbox{diag}(-,+,+,+)$.

\section{A new model of NED}

Let us consider NED with the Lagrangian density
\begin{equation}
{\cal L} = -\frac{{\cal F}}{(2\beta{\cal F}+1)^3}+\frac{\gamma}{2}{\cal G}^2,
\label{1}
\end{equation}
where $\beta$ ($\beta>0$) and $\gamma$ ($\gamma>0$) are dimensional parameters, ${\cal F}=(1/4)F_{\mu\nu}F^{\mu\nu}=(\textbf{B}^2-\textbf{E}^2)/2$, ${\cal G}=(1/4)F_{\mu\nu}\tilde{F}^{\mu\nu}=\textbf{B}\cdot\textbf{E}$ ($\tilde{F}^{\mu\nu}=\epsilon^{\mu\nu\alpha\beta}F_{\alpha\beta}/2$ is a dual tensor), $F_{\mu\nu}=\partial_\mu A_\nu-\partial_\nu A_\mu$ is the field strength tensor. The symmetrical energy-momentum tensor, found from Eq. (1), reads
\[
T_{\mu\nu}={\cal L}_{\cal F}F_\mu^{~\alpha}F_{\nu\alpha}+\frac{1}{2}{\cal L}_{\cal G}\left(F_\mu^{~\alpha}\tilde{F}_{\nu\alpha}
+F_\nu^{~\alpha}\tilde{F}_{\mu\alpha}\right)-g_{\mu\nu}{\cal L}
\]
\begin{equation}
=\frac{(4\beta{\cal F}-1)F_\mu^{~\alpha}F_{\nu\alpha}}{(1+2\beta{\cal F})^4}+\frac{1}{2}\gamma{\cal G}\left(F_\mu^\alpha\tilde{F}_{\nu\alpha}+F_\nu^{~\alpha}\tilde{F}_{\mu\alpha}\right)-g_{\mu\nu}{\cal L},
\label{2}
\end{equation}
where ${\cal L}_{\cal F}=\partial{\cal L}/\partial{\cal F}$, ${\cal L}_{\cal G}=\partial{\cal L}/\partial{\cal G}$.
From Eq. (2) we obtain the energy-momentum tensor trace
\begin{equation}
{\cal T}\equiv T_\mu^{~\mu}=\frac{24\beta{\cal F}^2}{(1+2\beta{\cal F})^4}+2\gamma {\cal G}^2.
\label{3}
\end{equation}
As $\beta=\gamma\rightarrow 0$ one arrives at Maxwell's electrodynamics, ${\cal L}\rightarrow -{\cal F}$ with the traceless energy-momentum tensor (${\cal T}\rightarrow 0$). The scale invariance of the model is broken because the energy-momentum tensor trace is not zero.
One can find the energy density $\rho$ and the pressure $p$ from Eq. (1)
\begin{equation}
\rho=-{\cal L}-E^2{\cal L}_{\cal F}+{\cal G}{\cal L}_{\cal G}=\frac{(1-4\beta{\cal F})E^2}{(1+2\beta{\cal F})^4} +\frac{{\cal F}}{(1+2\beta{\cal F})^3}+\frac{1}{2}\gamma{\cal G}^2,
\label{4}
\end{equation}
\begin{equation}
p={\cal L}+\frac{E^2-2B^2}{3}{\cal L}_{\cal F}-{\cal G}{\cal L}_{\cal G}=-\frac{{\cal F}}{(2\beta{\cal F}+1)^3}+\frac{(E^2-2B^2)(4\beta{\cal F}-1)}{3(2\beta{\cal F}+1)^4}-\frac{1}{2}\gamma{\cal G}^2.
\label{5}
\end{equation}
\subsection{The principles of causality and unitarity}

The causality principle requires that the group velocity of excitations over the background has to be less than the light speed. Then there are no tachyons in the spectrum of the theory. The absence of ghosts is guaranteed by the unitarity principle. Both principles hold if the inequalities are satisfied \cite{Shabad2}:
\[
 {\cal L}_{\cal F}\leq 0,~~~~{\cal L}_{{\cal F}{\cal F}}\geq 0,~~~~{\cal L}_{{\cal G}{\cal G}}\geq 0,
\] \begin{equation}
{\cal L}_{\cal F}+2{\cal F} {\cal L}_{{\cal F}{\cal F}}\leq 0,~~~~2{\cal F} {\cal L}_{{\cal G}{\cal G}}-{\cal L}_{\cal F}\geq 0.
\label{6}
\end{equation}
Making use of Eq. (1) we obtain
\[
{\cal L}_{\cal F}= \frac{4\beta{\cal F}-1}{(1+2\beta{\cal F})^4},~~~~ {\cal L}_{{\cal G}{\cal G}}=\gamma,~~~~
2{\cal F}{\cal L}_{{\cal G}{\cal G}}-{\cal L}_{{\cal F}}=2{\cal F}\gamma+\frac{1-4\beta{\cal F}}{(1+2\beta{\cal F})^4},
\]
\begin{equation}
{\cal L}_{\cal F}+2{\cal F}{\cal L}_{{\cal F}{\cal F}}=\frac{-40(\beta{\cal F})^2+26\beta{\cal F}-1}{(1+2\beta{\cal F})^5},~~~~
{\cal L}_{{\cal F}{\cal F}}=\frac{12\beta(1-2\beta{\cal F})}{(1+2\beta{\cal F})^5}.
\label{7}
\end{equation}
With the help of Eqs. (6) and (7), one finds that if $\gamma=0$ (${\cal L}_{\cal G}=0$) $\textbf{B}=0$, one obtains $|\textbf{E}|\leq \sqrt{1/\beta}$. But this restriction is satisfied because the maximum value of the electric field is $|\textbf{E}_{max}|= \sqrt{1/\beta}$ [see Eq. (19)].
When $\gamma=0$, $\textbf{E}=0$, we have $|\textbf{B}|\leq \sqrt{(13-\sqrt{129})/(20\beta)}\approx 0.2865/\sqrt{\beta}$.

\section{Field equations}

Making use of Eq. (1), we obtain equations of motion
\begin{equation}
\partial_\mu\left({\cal L}_{\cal F}F^{\mu\nu} +{\cal L}_{\cal G}\tilde{F}^{\mu\nu} \right)=0.
\label{8}
\end{equation}
With the help of Eqs. (1) and (8), one finds
\begin{equation}
 \partial_\mu\left(\frac{(4\beta{\cal F}-1)F^{\mu\nu}}{(1+2\beta{\cal F})^4}
+\gamma{\cal G}\tilde{F}^{\mu\nu}\right)=0.
\label{9}
\end{equation}
The electric displacement field is defined as $\textbf{D}=\partial{\cal L}/\partial \textbf{E}$,
\begin{equation}
\textbf{D}=\frac{1-4\beta{\cal F}}{(1+2\beta{\cal F})^4} \textbf{E}+\gamma {\cal G}\textbf{B}.
\label{10}
\end{equation}
We obtain the magnetic field $\textbf{H}=-\partial{\cal L}/\partial \textbf{B}$,
\begin{equation}
\textbf{H}= \frac{1-4\beta{\cal F}}{(1+2\beta{\cal F})^4}\textbf{B}-\gamma{\cal G}\textbf{E}.
\label{11}
\end{equation}
It is convenient to use the decomposition of Eqs. (10) and (11) as \cite{Hehl}
\begin{equation}
D_i=\varepsilon_{ij}E^j+\nu_{ij}B^j,~~~~H_i=(\mu^{-1})_{ij}B^j-\nu_{ji}E^j.
\label{12}
\end{equation}
Making use of Eqs. (10), (11) and (12), one finds
\[
\varepsilon_{ij}=\delta_{ij}\varepsilon,~~~~(\mu^{-1})_{ij}=\delta_{ij}\mu^{-1},~~~~\nu_{ji}=\delta_{ij}\nu,
\]
\begin{equation}
\varepsilon=\frac{1-4\beta{\cal F}}{(1+2\beta{\cal F})^4},~~~~
\mu^{-1}=\varepsilon=\frac{1-4\beta{\cal F}}{(1+2\beta{\cal F})^4},~~~~\nu=\gamma {\cal G}.
\label{13}
\end{equation}
By virtue of Eqs. (10) and (11), field equations (9) can be written as Maxwell's equations
\begin{equation}
\nabla\cdot \textbf{D}= 0,~~~~ \frac{\partial\textbf{D}}{\partial
t}-\nabla\times\textbf{H}=0.
\label{14}
\end{equation}
Equation (14) are represented as nonlinear Maxwell's equations because $\varepsilon_{ij}$, $(\mu^{-1})_{ij}$, and $\nu_{ji}$ depend on electromagnetic fields.
With the aid of the Bianchi identity $\partial_\mu \tilde{F}^{\mu\nu}=0$, we obtain the second pair of Maxwell's equations
\begin{equation}
\nabla\cdot \textbf{B}= 0,~~~~ \frac{\partial\textbf{B}}{\partial
t}+\nabla\times\textbf{E}=0.
\label{15}
\end{equation}
Making use Eqs. (10) and (11), one finds the equation as follows:
\begin{equation}
\textbf{D}\cdot\textbf{H}=(\varepsilon^2-\nu^2)\textbf{E}\cdot\textbf{B}+2\varepsilon\nu{\cal F}.
\label{16}
\end{equation}
The dual symmetry is broken in our model because $\textbf{D}\cdot\textbf{H}\neq\textbf{E}\cdot\textbf{B}$  \cite{Gibbons}. In Maxwell's electrodynamics ($\varepsilon=1$, $\nu=0$) and in BI electrodynamics the dual symmetry holds, but in QED with quantum corrections and in generalized BI electrodynamics \cite{Krug} the dual symmetry is broken.

\subsection{The fields of point-like electric charges}

The electric displacement field for point-like particle with the electric charge $q_e$ obeys the equation
\begin{equation}
\nabla\cdot \textbf{D}=4\pi q_e\delta(\textbf{r}).
\label{17}
\end{equation}
The solution to Eq. (17), making use of Eq. (10), at $\textbf{B}=0$ reads
\begin{equation}
\frac{E\left(1+2\beta E^2\right)}{(1-\beta E^2)^4}=\frac{q_e}{r^2}.
\label{18}
\end{equation}
If $r\rightarrow 0$ the solution, in accordance with Eq. (18), is given by
\begin{equation}
E(0)=\sqrt{\frac{1}{\beta}}.
\label{19}
\end{equation}
In the center of the point-like charges, the singularity of the electric field is absent, but in Maxwell's electrodynamics the singularity in the center of charges holds.
The maximum of the electric field at the origin of charged particles is given by Eq. (19). In BI electrodynamics also there is no singularity in the center of charges. It is convenient to introduce unitless variables
\begin{equation}
x=\frac{r^2}{q_e\sqrt{\beta}},~~~~y=\sqrt{\beta}E.
\label{20}
\end{equation}
Then Eq. (18) can be written as follows:
\begin{equation}
\frac{y(1+2y^2)}{(1-y^2)^4}=\frac{1}{x}.
\label{21}
\end{equation}
The plot of the function $y(x)$ is given in Fig. 1.
\begin{figure}[h]
\includegraphics{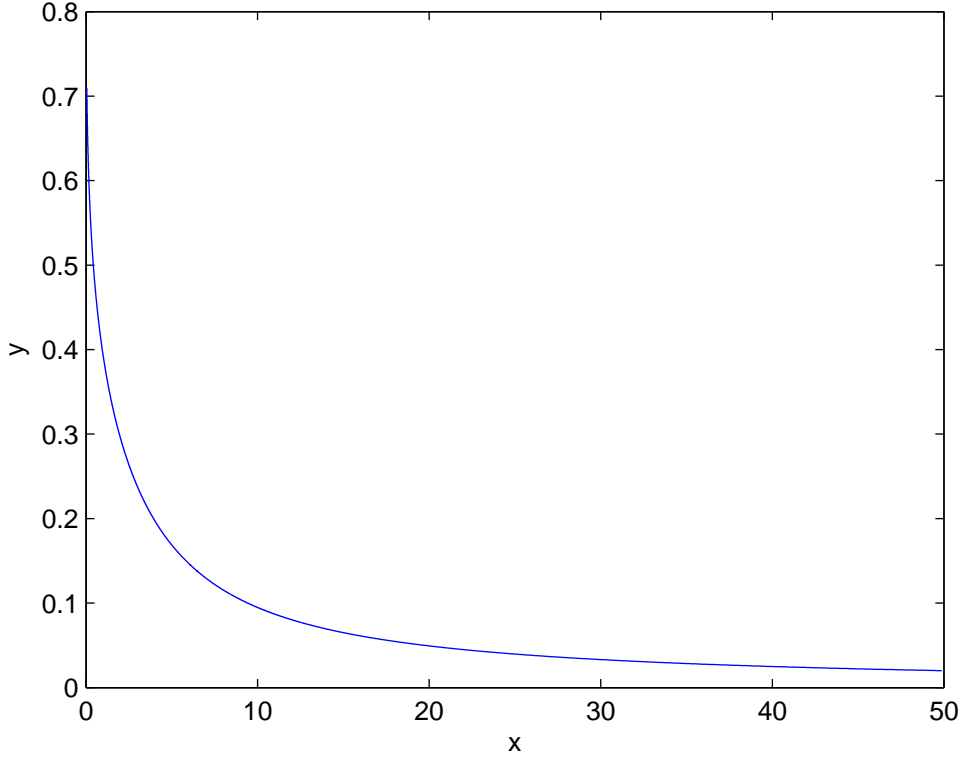}
\caption{\label{fig.1}The function  $y$ versus $x$.}
\end{figure}
The approximate real and positive solutions to Eq. (21) are in Table 1.
\begin{table}[ht]
\caption{}
\centering
\begin{tabular}{c c c c c c c c c c  c}\\[1ex]
\hline
$x$ & 1 & 2 & 3 & 4 & 5 & 6 & 7 & 8 & 9 & 10\\[0.5ex]
\hline
 $y$ & 0.392 & 0.295 & 0.237 & 0.198 & 0.169 & 0.147 & 0.129 & 0.115 & 0.104 & 0.095\\[0.5ex]
\hline
\end{tabular}
\end{table}
The function $y(x)$ as $x\rightarrow \infty$ is
\begin{equation}
y=\frac{1}{x}+{\cal O}(x^{-2}).
\label{22}
\end{equation}
  The asymptotic of the electric field as $r\rightarrow\infty$, using Eqs. (20) and (22), leads to
\begin{equation}
E(r)=\frac{q_e}{r^2}+{\cal O}(r^{-4}).
\label{23}
\end{equation}
Equation (23) shows corrections to Coulomb's law. In classical electrodynamics
at $\beta=0$ one has the Coulomb law $E=q_e/r^2$.
Making use of Eq. (21), we find the asymptotic of $y$ as $x\rightarrow 0$
\begin{equation}
y(x)=1-\frac{\sqrt[4]{3x}}{2}~~~~~~~x\rightarrow 0.
\label{24}
\end{equation}
Then from Eq. (20), one obtains
\begin{equation}
E(r)=\frac{1}{\sqrt{\beta}}-\frac{\sqrt[4]{3}\sqrt{r}}{2\sqrt[4]{q_e}\beta^{5/8}}~~~~~~r\rightarrow 0.
\label{25}
\end{equation}
At $r=0$ we come to Eq. (19). Equation (25) shows the behavior of the electric field over short distances.

\section{Cosmology}

The action of GR, where electromagnetic fields are the source of gravitational fields, is given by
\begin{equation}
S=\int d^4x\sqrt{-g}\left[\frac{1}{2\kappa^2}R+ {\cal L}\right],
\label{26}
\end{equation}
where $R$ is the Ricci scalar and $M_{Pl}=\kappa^{-1}$ is the reduced Planck mass. Varying action (26), we obtain the Einstein and electromagnetic field equations
\begin{equation}
R_{\mu\nu}-\frac{1}{2}g_{\mu\nu}R=-\kappa^2T_{\mu\nu},
\label{27}
\end{equation}
\begin{equation}
\partial_\mu\left(\frac{\sqrt{-g}F^{\mu\nu}(4\beta{\cal F}-1)}{(2\beta{\cal F}+1)^4}\right)=0.
\label{28}
\end{equation}
The line element of homogeneous and isotropic cosmological spacetime is
\begin{equation}
ds^2=-dt^2+a(t)^2\left(dx^2+dy^2+dz^2\right),
\label{29}
\end{equation}
where $a(t)$ is a scale factor. We suppose that the cosmic background is stochastic magnetic fields. Averaging the magnetic fields, that are sources of gravitational fields \cite{Tolman}, one has the isotropy of the Friedman$-$Robertson$-$ Walker (FRW) spacetime.
The magnetic fields averaged obey equations as follows:
\begin{equation}
\langle \textbf{B}\rangle=0,~~~~\langle E_iB_j\rangle=0,~~~~\langle B_iB_j\rangle=\frac{1}{3}B^2g_{ij},
\label{30}
\end{equation}
where the brackets $\langle.\rangle$ denote an average over a volume. In the following we omit the brackets $\langle.\rangle$. The energy-momentum tensor of NED with Eqs. (30) can be represented as a perfect fluid \cite{Novello1}.
The Friedmann equation is written as
\begin{equation}
3\frac{\ddot{a}}{a}=-\frac{\kappa^2}{2}\left(\rho+3p\right),
\label{31}
\end{equation}
where the dots over the letter mean the derivatives with respect to the cosmic time. If $\rho + 3p < 0$ the acceleration of the universe occurs. In accordance with the standard cosmological model, there is an isotropic symmetry, and $\langle B_i\rangle = 0$. With the help of Eqs. (4) and (5) one obtains (if $\textbf{E}=0$)
\begin{equation}
\rho+3p=-\frac{B^2(5\beta B^2-1)}{(1+\beta B^2)^4}.
\label{32}
\end{equation}
The plot of the function $\beta(\rho+3p)$ versus $\beta B^2$ is depicted in Fig. 2.
\begin{figure}[h]
\includegraphics{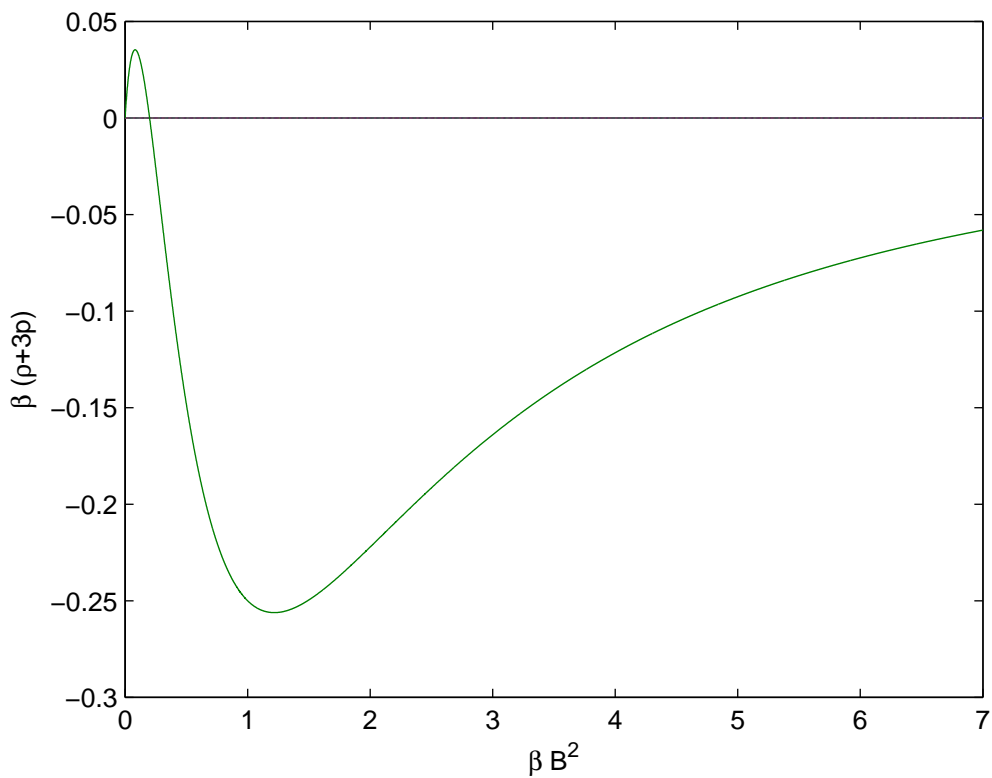}
\caption{\label{fig.2}The function  $\beta(\rho+3p)$ versus $\beta B^2$. }
\end{figure}
When $\rho + 3p < 0$ the acceleration of the universe holds and $\beta B^2>1/5$. Thus, the strong magnetic fields result in the universe inflation. The conservation of the energy-momentum tensor, $\nabla^\mu T_{\mu\nu}=0$, gives the equation
\begin{equation}
\dot{\rho}+3\frac{\dot{a}}{a}\left(\rho+p\right)=0.
\label{33}
\end{equation}
Making use of Eqs. (4) and (5), in the case $\textbf{E} = 0$, we obtain
\begin{equation}
\rho=\frac{B^2}{2\left(1+\beta B^2\right)^3},~~~~\rho+p=\frac{2B^2(1-2\beta B^2)}{3\left(1+\beta B^2\right)^4}.
\label{34}
\end{equation}
Integrating Eq. (33), using Eq. (34), one finds
\begin{equation}
B(t)=\frac{B_0}{a^2(t)},
\label{35}
\end{equation}
were $B_0$ is the magnetic field corresponding to the value $a(t)=1$.
As a result, the scale factor increases because of inflation, and the magnetic field decreases. Making use of Eqs. (34) and (35), we obtain the energy density and pressure
\begin{equation}
\rho(t)=\frac{a^8(t) B_0^2}{2\left(a^4(t)+\beta B_0^2\right)^3},
~~~~p(t)=\frac{a^8(t)B_0^2(a^4(t)-11\beta B_0^2)}{6\left(a^4(t)+\beta B_0^2\right)^4}.
\label{36}
\end{equation}
From Eq. (36) one finds
\begin{equation}
\lim_{a(t)\rightarrow 0}\rho(t)=\lim_{a(t)\rightarrow 0}p(t)=\lim_{a(t)\rightarrow \infty}\rho(t)=\lim_{a(t)\rightarrow \infty}p(t)=0.
\label{37}
\end{equation}
As a result, singularities of the energy density and pressure as $a(t)\rightarrow 0$ and $a(t)\rightarrow \infty$ are absent. The equation of state (EoS) $w=p(t)/\rho(t)$ versus $x=a(t)/(\beta B_0^2)^{1/4}$ is depicted in Fig. 3.
\begin{figure}[h]
\includegraphics{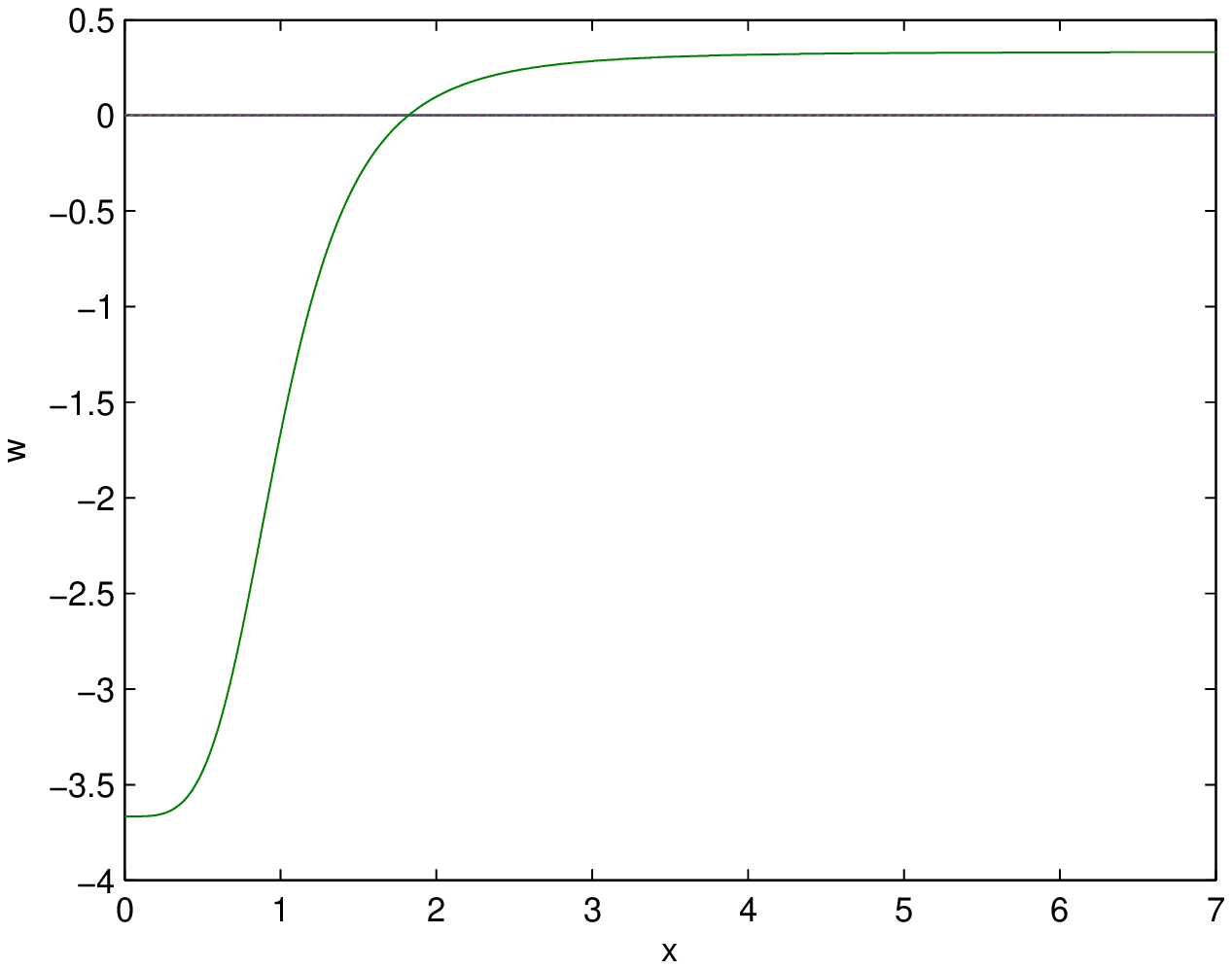}
\caption{\label{fig.3}The function  $w$ versus $x=a/(\beta B_0^2)^{1/4}$.}
\end{figure}
With the help of Eq. (36) we obtain
\begin{equation}
\lim_{x\rightarrow\infty} w=\lim_{x\rightarrow\infty}\frac{x^4-11}{3(x^4+1)}=\frac{1}{3}.
\label{38}
\end{equation}
As $a(t)\rightarrow \infty$ one has the EoS for ultra-relativistic case \cite{Landau}.
For $x=\sqrt[4]{2}\approx 1.19$, EoS corresponds to de Sitter spacetime, $w=-1$.
 From the Einstein equation (27) and Eq. (3), we obtain the curvature
\begin{equation}
R=\kappa^2{\cal T}=\frac{6\kappa^2\beta B^4}{(1+\beta B^2)^4}=\kappa^2\left[\rho(t)-3p(t)\right].
\label{39}
\end{equation}
The plot of the function $\beta R/\kappa^2$ versus $[1/(\beta B_0^2)]^{1/4}a$ is depicted in Fig. 4.
\begin{figure}[h]
\includegraphics{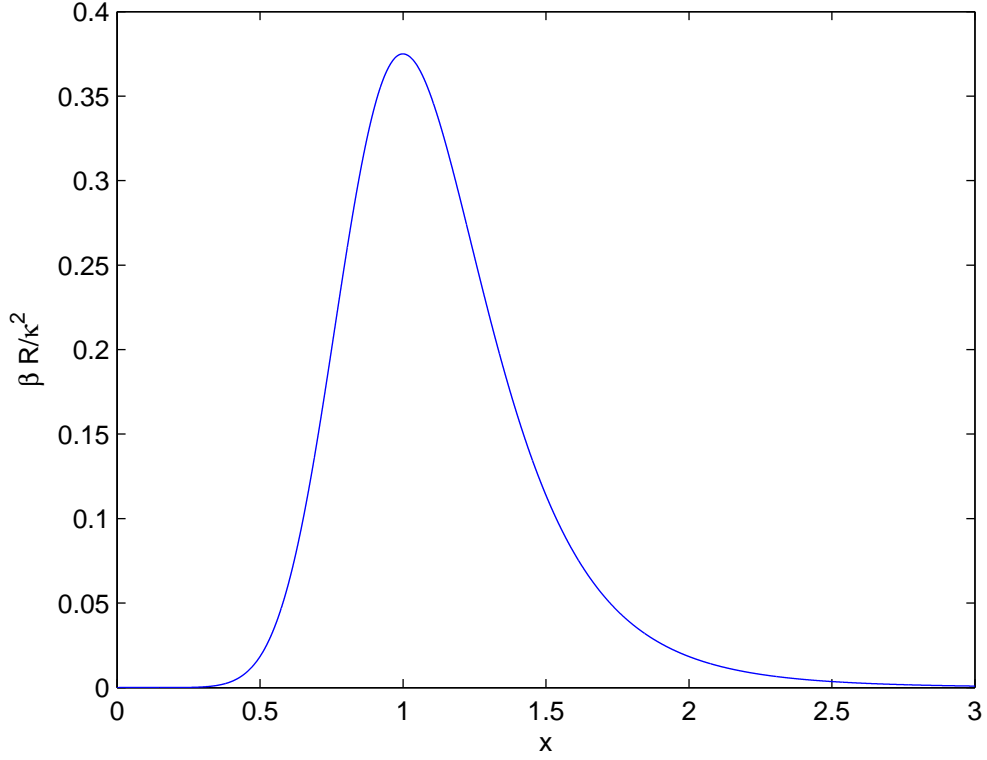}
\caption{\label{fig.4}The function  $\beta R/\kappa^2$ versus $x\equiv a/(\beta B_0^2)^{1/4}$. }
\end{figure}
Making use of Eqs. (37) and (39), we find
\begin{equation}
\lim_{a(t)\rightarrow 0}R(t)=\lim_{a(t)\rightarrow \infty}R(t)=0.
\label{40}
\end{equation}
The singularity of the Ricci scalar is absent. The Ricci tensor squared $R_{\mu\nu}R^{\mu\nu}$ and the Kretschmann scalar $R_{\mu\nu\alpha\beta}R^{\mu\nu\alpha\beta}$ may be expressed as linear combinations of $\kappa^4\rho^2$, $\kappa^4\rho p$, and $\kappa^4p^2$ \cite{Kruglov1} and, according to Eq. (37), they vanish at $a(t)\rightarrow 0$ and $a(t)\rightarrow \infty$.
As $t\rightarrow\infty$ the scale factor increases and spacetime approaches to the Minkowski spacetime. From Eqs. (32) and (35) we find that the universe accelerates at $a(t)<(5\beta)^{1/4}\sqrt{B_0}\approx 1.5\beta^{1/4}\sqrt{B_0}$ and the universe inflation occurs.

\section{ Evolution of the universe}

The second Friedmann equation for three dimensional flat universe is given by
\begin{equation}
\left(\frac{\dot{a}}{a}\right)^2=\frac{\kappa^2\rho}{3}.
\label{41}
\end{equation}
With the help of Eqs. (34) and (35), and making use of Eq. (41), we obtain
\begin{equation}
\dot{a} =\frac{\kappa B_0a^5}{\sqrt{6}(a^4+\beta B_0^2)^{3/2}}.
\label{42}
\end{equation}
Making use of the unitless variable $x=a/(\beta^{1/4}\sqrt{B_0})$, Eq. (42) becomes
\begin{equation}
\dot{x} =\frac{\kappa x^5}{\sqrt{6\beta}(x^4+1)^{3/2}}.
\label{43}
\end{equation}
The plot of the function $y\equiv\sqrt{6\beta}\dot{x}/\kappa$ versus $x$ is depicted in Fig. 5.
\begin{figure}[h]
\includegraphics{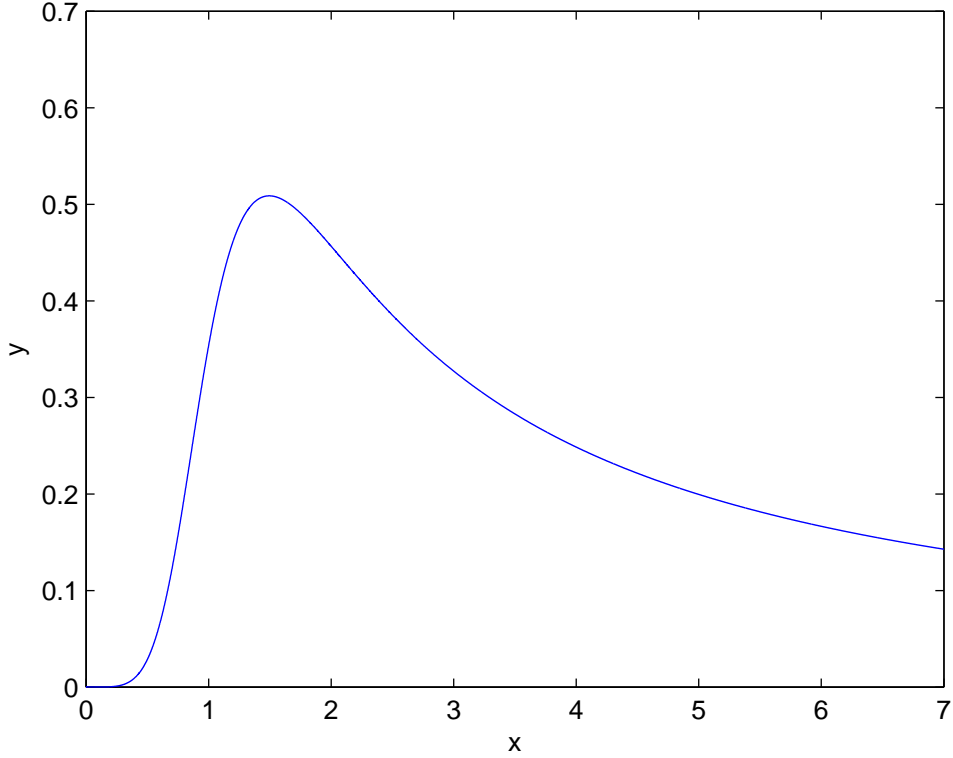}
\caption{\label{fig.5}The function $y\equiv\sqrt{6\beta}\dot{x}/\kappa$ versus $x$.}
\end{figure}
Figure shows that at the initial time the universe accelerates ($\dot{y}>0$) till the graceful exit point $x=\sqrt[4]{5}$ ($\dot{y}=0$) at the inflation end. Then the universe decelerates.
Integrating Eq. (43)
\begin{equation}
\int_\epsilon^x \frac{(x^4+1)^{3/2}}{x^5}dx =\frac{\kappa}{\sqrt{6\beta}}\int_0^t dt,
\label{44}
\end{equation}
we arrive at the equation
\[
0.5\left(\sqrt{x^4+1}-\sqrt{\epsilon^4+1}\right)-0.25\left(\frac{\sqrt{x^4+1}}{x^4}
-\frac{\sqrt{\epsilon^4+1}}{\epsilon^4}\right)
\]
\begin{equation}
-0.75\tanh^{-1}\left(\frac{\sqrt{x^4+1}-\sqrt{\epsilon^4+1}}{1-\sqrt{(x^4+1)(\epsilon^4+1)}}\right)=\frac{\kappa t}{\sqrt{6\beta}},
\label{45}
\end{equation}
where $\tanh^{-1}(x)$ is the inverse $\tanh$-function, $\epsilon$ corresponds to the beginning of the universe inflation. Equation (45) allows us to study the evolution of the universe inflation.
The analytical solution to Eq. (45) is unknown.
The scale factor as $t\rightarrow\infty$, in the leading order, leads to
\begin{equation}
a(t)=\sqrt[4]{\frac{2}{3}}\sqrt{\kappa B_0 t}\approx 0.9\sqrt{\kappa B_0 t},
\label{46}
\end{equation}
and corresponds to the radiation era.
To describe the expansion of the universe, we introduce the deceleration parameter, making use of Eqs. (31), (34), (35) and (41),
\begin{equation}
q=-\frac{\ddot{a}a}{(\dot{a})^2}=\frac{x^4-5}{x^4+1}.
\label{47}
\end{equation}
\begin{figure}[h]
\includegraphics{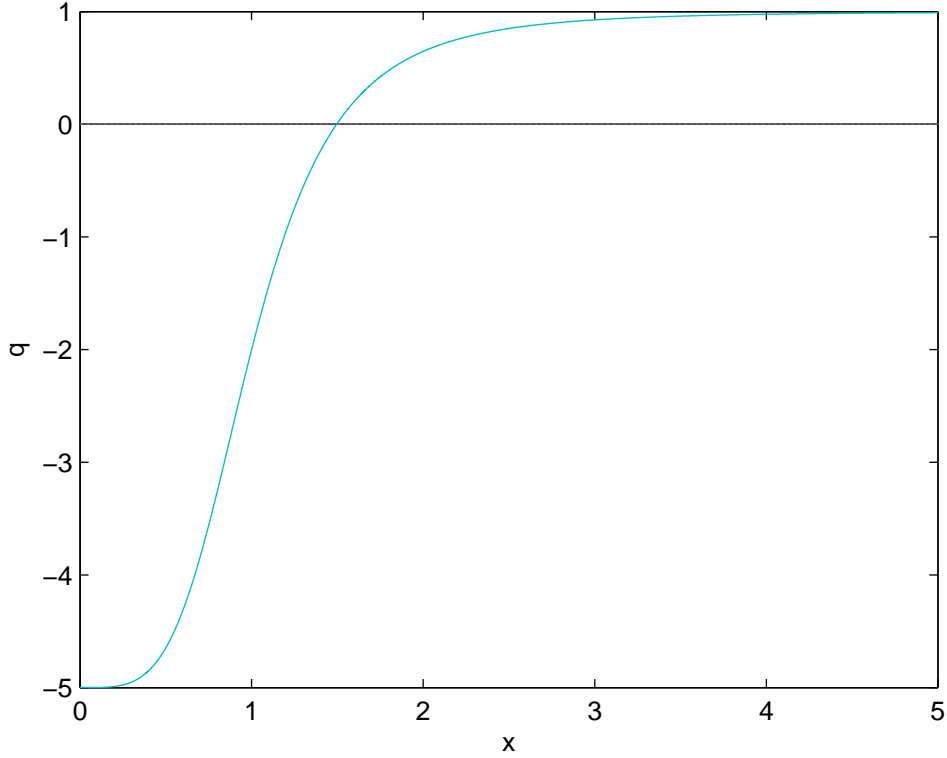}
\caption{\label{fig.6}The function $q$ versus $x=a/(\beta B_0^2)^{1/4}$.}
\end{figure}
Figure 6 shows the behavior of the deceleration parameter $q$ versus $x=a/(\beta B_0^2)^{1/4}$. The inflation ($q<0$ ) lasts till the graceful exit $x=\sqrt[4]{5}\approx 1.5$.
At $x=\sqrt[4]{5}$ the deceleration parameter becomes zero and after the deceleration phase ($q>0$) occurs. 
The similar behavior of the scale factor takes place in another model proposed in \cite{Kruglov4}.

To estimate the amount of the inflation, we use the definition of e-foldings \cite{Liddle}
\begin{equation}
N=\ln\frac{a(t_{end})}{a(t_{in})},
\label{48}
\end{equation}
where $t_{end}$ is the final time of the inflation and $t_{in}$ is an initial time. Taking into account the graceful exit point $x\approx 1.5$ we obtain $a(t_{end})\approx 1.5 b$ ($b\equiv \beta^{1/4}\sqrt{B_0}$).
The horizon and flatness problems can be solved if e-foldings $N\approx 70$ \cite{Liddle}. Then one obtains from Eq. (48) the scale factor corresponding to the initial time of the inflation
\begin{equation}
a(t_{in})=\frac{1.5b}{\exp(70)}\approx 6\times 10^{-31}b,
\label{49}
\end{equation}
and $\epsilon\approx 6\times 10^{-31}$.
To estimate the duration of the inflationary period we should analyze Eq. (45). Making use of particle physics units ($c=\hbar=1$) $\kappa=\sqrt{8\pi G}=4.1\times 10^{-28}~\mbox{eV}^{-1}$, $\beta=2.3\times 10^{-29}~\mbox{eV}^{-4}$ (see ``Appendix"), $1~\mbox{s}=1.5\times 10^{15}~\mbox{eV}^{-1}$, we obtain $\kappa/\sqrt{6\beta}=0.35\times10^{-13}\mbox{eV}=52.5~\mbox{s}^{-1}$.
 If one uses the value $x=\sqrt[4]{5}\approx 1.5$ corresponding to the end of the inflation, the result for the duration of the universe inflation, following from Eq. (45), will be huge. Thus, the  universe inflation will be almost eternal, i.e., the current universe acceleration can be explained. But in this case it is questionable to describe nucleosynthesis and other epochs. When one takes the time duration $1~\mbox{s}$ Eq. (44) gives for $x\approx 1.5$ the value $\epsilon=0.2665$. But in this case the e-folding number (48) is $N\approx 1.7$ which is too small to explain the horizon and flatness problems. As a result, the model proposed probably cannot be used to solve all problems in the early and the late time universe acceleration and can be considered as a toy-model. By varying the parameter $\epsilon$ of the initial time, one can analyze different scenarios of the universe acceleration, the e-foldings number, and the duration of the inflation. The attractive feature of this model is the existence of the phases of the universe acceleration, deceleration and the graceful exit.

\subsection{Speed of sound and causality}

When the speed of the sound is less than the local light speed, $c_s\leq 1$ \cite{Quiros} the causality holds. A classical stability is guarantied if the square sound speed is positive, $c^2_s> 0$. Making use of Eqs. (4) and (5), we obtain the sound speed squared ($E=0$)
\begin{equation}
c^2_s=\frac{dp}{d\rho}=\frac{dp/d{\cal F}}{d\rho/d{\cal F}}=\frac{22\beta^2 B^4-25\beta B^2+1)}{3(\beta B^2+1)(1-2\beta B^2)}.
\label{50}
\end{equation}
A requirement of the classical stability ($c^2_s> 0$) results
\begin{equation}
0<\beta B^2<\frac{25-\sqrt{537}}{44}\approx 0.04~~\mbox{or}~~0.5<\beta B^2<\frac{25+\sqrt{537}}{44}\approx 1.09.
\label{51}
\end{equation}
The causality ($c^2_s\leq 1$) leads to
\begin{equation}
0\leq\beta B^2\leq 0.5~~\mbox{or}~~\beta B^2\geq \frac{11+\sqrt{177}}{28}\approx 0.87.
\label{52}
\end{equation}
Both Eqs. (51) and (52) lead to $0\leq\beta B^2<(25-\sqrt{537})/44$ or $(11+\sqrt{177})/28\leq\beta B^2<(25+\sqrt{537})/44$.
These requirements and the principles of causality and unitarity, studied in Sect. 2, hold when the inequality $0\leq\beta B^2<(25-\sqrt{537})/44\approx 0.04$ holds at the deceleration phase of the universe evolution. At the acceleration phase,
at $\beta B^2>0.2$, the classical stability, causality and unitarity are broken.

\section{Cosmological parameters}

Making use of Eqs. (4) and (5) we obtain (at $\textbf{E}=0$)
\begin{equation}
p=-\rho+\frac{4\rho(1-2\beta B^2)}{3(\beta B^2+1)},
\label{53}
\end{equation}
\begin{equation}\label{54}
 2\rho\beta (\beta B^2+1)^3-\beta B^2=0.
\end{equation}
By introducing the variable $z=\beta B^2+1$, from Eq. (54), we find the cubic equation
\begin{equation}\label{55}
 2\rho\beta z^3-z+1=0.
\end{equation}
The determinant of Eq. (55) is positive, and therefore, there is one real solution and two nonphysical solutions. The real solution reads
\begin{equation}\label{56}
z=-\frac{2}{\sqrt{6\beta \rho}}\cosh\left(\frac{\varphi}{3}\right), ~~~\cosh(\varphi)=\sqrt{\frac{27\beta\rho}{2}},
\end{equation}
which leads to the solution
\begin{equation}\label{57}
\beta B^2=-1-\frac{2}{\sqrt{6\beta \rho}}\cosh\left[\frac{1}{3}\ln\left(\sqrt{\frac{27\beta\rho}{2}}+\sqrt{\frac{27\beta\rho}{2}-1}\right)\right].
\end{equation}
Making use of Eqs. (53) and (57), we obtain EoS for the perfect fluid
\begin{equation}
p=-\rho+f(\rho),~~f(\rho)=-\frac{2\rho\left(3\sqrt{6\beta\rho}+4\cosh\left[\frac{1}{3}\ln\left(\sqrt{\frac{27\beta\rho}{2}}
+\sqrt{\frac{27\beta\rho}{2}-1}\right)\right]\right)}{3\cosh\left[\frac{1}{3}\ln\left(\sqrt{\frac{27\beta\rho}{2}}
+\sqrt{\frac{27\beta\rho}{2}-1}\right)\right]}.
\label{58}
\end{equation}
The plot of the function $p\beta$ versus $\rho\beta$ is depicted in Fig. 7.
\begin{figure}[h]
\includegraphics{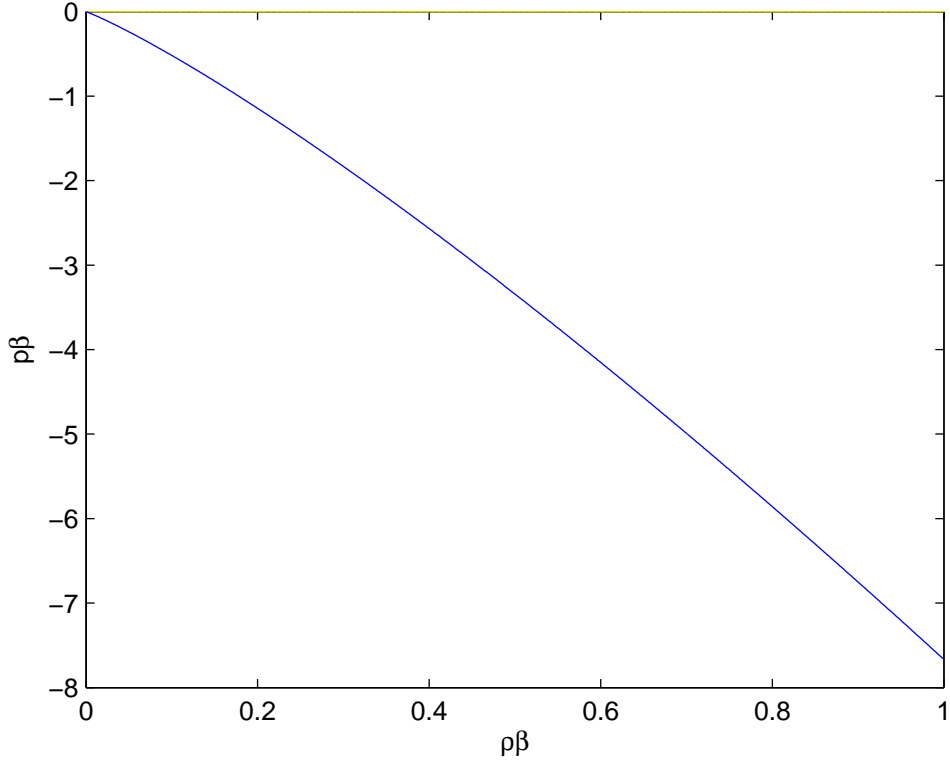}
\caption{\label{fig.7}The function $p\beta$ versus $\rho\beta$.}
\end{figure}
If the condition $|f(\rho)/\rho|\ll 1$ occurs during the inflation, the expressions for the spectral index $n_s$, the tensor-to-scalar ratio $r$, and the running of the spectral index $\alpha_s=dn_s/d\ln k$ are given by \cite{Odintsov}
\begin{equation}
n_s\approx 1-6\frac{f(\rho)}{\rho},~~~r\approx 24\frac{f(\rho)}{\rho},~~~\alpha_s\approx -9\left(\frac{f(\rho)}{\rho}\right)^2.
\label{59}
\end{equation}
 Equation (59) lead to the relations
 \[
r=4(1-n_s)=8\sqrt{-\alpha_s}
 \]
\begin{equation}
=-\frac{16\left(3\sqrt{6\beta\rho}+4\cosh\left[\frac{1}{3}\ln\left(\sqrt{\frac{27\beta\rho}{2}}
+\sqrt{\frac{27\beta\rho}{2}-1}\right)\right]\right)}{\cosh\left[\frac{1}{3}\ln\left(\sqrt{\frac{27\beta\rho}{2}}
+\sqrt{\frac{27\beta\rho}{2}-1}\right)\right]}.
\label{60}
\end{equation}
The PLANCK experiment \cite{Ade} and WMAP data \cite{Komatsu}, \cite{Hinshaw} result
\[
n_s=0.9603\pm 0.0073 ~(68\% CL),~~~r<0.11 ~(95\%CL),
\]
\begin{equation}
\alpha_s=-0.0134\pm0.0090 ~(68\% CL).
\label{61}
\end{equation}
If $r=0.13$ we obtain from Eqs. (60) the values for the spectral index $n_s=0.9675$ and the running of the spectral index $\alpha_s=-2.64\times 10^{-4}$. Making use of Eq. (60), one finds the value $\beta\rho\approx 0.074$ corresponding to values of cosmological parameters which gives the value of the magnetic field $B\approx 0.7/\sqrt{\beta}\approx 146~\mbox{MeV}^2$ (see ``Appendix") and corresponds to the inflation phase. It is worth noting that the maximum of the energy density occurs at $\rho_{max}=2/(27\beta)\approx 0.07407/\beta$.

\section{Conclusion}

A new model of NED, without a singularity of the electric field in the center of charges, has been proposed.
The principles of causality and unitarity were studied. We found the range of electromagnetic fields when causality and unitarity hold. It was demonstrated that the dual symmetry is broken in our model. We obtained corrections to Coulomb's law that are in the order of ${\cal O}(r^{-4})$. Gravitational field coupled with our NED was studied. We considered the magnetic universe with a stochastic background, $\langle B^2\rangle\neq 0$.
It was demonstrated that the model with homogeneous and isotropic cosmology describes the universe inflation. The singularities of the energy density, pressure, the Ricci scalar, the Ricci tensor squared, and the Kretschmann scalar are absent. We shown that a stochastic magnetic field, in the framework of NED proposed, is the source of the universe inflation at the early epoch. Then as $B< 1/\sqrt{5\beta}$, the universe decelerates approaching to the radiation era.
 The interval of magnetic fields when the classical stability and the causality occur was obtained.
We calculated the spectral index, the tensor-to-scalar ratio, and the running of the spectral index which are in approximate agreement with the PLANK and WMAP data.
It worth noting that in our model of inflation the graceful exit problem is absent. Figures 2 and 5 show that after inflation the universe decelerates approaching to the radiation era.
But the problem of the current acceleration exists. One may solve this problem by
modifying general relativity. Another way is to introduce scalar fields coupled nonminimally with
gravity. For classical electrodynamics coupled with $F(R)$ gravity,
this was studied in \cite{Odintsov1}. The problem of late-time acceleration of the universe is long standing problem that can be solved by introduction of the cosmological constant. One can generalize the model under consideration to consider nonminimal coupling gravity with NED.

\section{Appendix}

It is natural that our model at weak electromagnetic fields will be converted into QED with loop corrections. Here, we define the model parameters $\beta$ and $\gamma$ by comparing (1), at weak field limit, with the Heisenberg$-$Euler Lagrangian. At $\beta{\cal F}\ll 1$ Lagrangian (1) becomes
\begin{equation}
{\cal L}=-{\cal F}+6\beta{\cal F}^2-24\beta^2{\cal F}^3+{\cal O}\left((\beta{\cal F})^4\right)+\frac{\gamma}{2}{\cal G}^2.
\label{62}
\end{equation}
The lowest order of the Heisenberg$-$Euler Lagrangian (QED with one loop correction) is given by \cite{Gies}
\begin{equation}
{\cal L}_{HE}=-{\cal F}+c_1{\cal F}^2+c_2{\cal G}^2,~~~c_2=\frac{14\alpha^2}{45m_e^4},~~~c_1=\frac{8\alpha^2}{45m_e^4},
\label{63}
\end{equation}
where the coupling constant $\alpha=e^2/(4\pi)\approx 1/137$ and the electron mass $m_e=0.51~\mbox{MeV}$. Comparing Eqs. (62) and (63) we obtain
\begin{equation}
\beta=\frac{4\alpha^2}{135m_e^4}=2.3\times 10^{-5}~\mbox{MeV}^{-4},~~~\gamma=\frac{28\alpha^2}{45m_e^4}= 4.9\times 10^{-4}~\mbox{MeV}^{-4}.
\label{64}
\end{equation}

\end{document}